\documentclass[12pt]{article}
\usepackage{amsmath,amssymb,exscale}
\usepackage{epsfig,psfrag}
\usepackage{color,amsmath,amscd,shadow,fancybox,axodraw,booktabs}
\input epsf
\newcommand{\gslashkk}{{\not \!G}^{(1)}}

\newcommand{\be}{\begin{equation}}
\newcommand{\ee}{\end{equation}}
\newcommand{\bear}{\begin{eqnarray}}
\newcommand{\eear}{\end{eqnarray}}
\newcommand{\ba}{\begin{array}}
\newcommand{\ea}{\end{array}}
\newcommand{\lae}{\begin{array}{c}\,\sim\vspace{-21pt}\\<
\end{array}}

\begin{document}
\topmargin -1.0cm
\oddsidemargin -0.8cm
\evensidemargin -0.8cm

\thispagestyle{empty}
\vspace{20pt}

\begin{center}

{\Large \bf
A Strongly Coupled Fourth Generation at the LHC }

\end{center}

\vspace{15pt}
\begin{center}
{\large  
G. Burdman, L. Da Rold, O.~J.~P.~\'Eboli and R. D. Matheus
} 

\vspace{20pt}
\textit{Departamento de F\'{i}sica Matem\'{a}tica\\ 
Instituto de F\'{i}sica, Universidade de S\~{a}o Paulo,
\\ R. do Mat\~{a}o 187, S\~{a}o Paulo, SP 05508-900, Brazil
}

\end{center}

\vspace{20pt}
\begin{center}
\textbf{Abstract}
\end{center}
\vspace{5pt} {\small \noindent

  We study extensions of the standard model with a strongly coupled
  fourth generation.  This occurs in models where electroweak
  symmetry breaking is triggered by the condensation of at least some
  of the fourth-generation fermions.  With focus on the phenomenology
  at the LHC, we study the pair production of fourth-generation down
  quarks, $D_4$. We consider the typical masses that could be
  associated with a strongly coupled fermion sector, in the 
  range (300--600)~GeV.  We show that the production and successive
  decay of these heavy quarks into final states with same-sign
  dileptons, trileptons and four leptons, can be easily seen above
  background with  relatively low luminosity. On the other hand, 
  in order to confirm the presence of a new strong interaction
  responsible for fourth-generation condensation, we study 
  its contribution to $D_4$ pair-production, and the potential to
  separate it from standard QCD-induced heavy quark production. 
  We show that this separation might require large amounts of
data. This is true even if it is assumed that the new interaction
is mediated by a massive colored vector boson, since its strong
coupling to the fourth generation renders its width of the order of its
mass. We conclude that, although  this class of models can be 
falsified at early stages of the LHC running, its confirmation would require
high integrated luminosities.
}


\vfill


\section{Introduction}

The origin of electroweak symmetry breaking (EWSB) is 
one of the most important questions in particle physics today. 
A natural solution to the quantum instability of the Higgs potential suggests that 
there should be new physics at the TeV scale.  
The central task of the
Large Hadron Collider (LHC) is to illuminate this question. 
An appealing and economic mechanism to explain EWSB 
is the condensation of the top quark, leading to a
unified description of the mechanism of symmetry breaking and 
the top mass~\cite{Bardeen:1989ds}. However, if this scenario
is to solve the hierarchy problem, the top quark should be
considerably heavier, around $m_t\sim (600-700)~$GeV. 
Already in Ref.~\cite{Bardeen:1989ds} it was suggested that the simplest model could be
extended to the condensation of a heavy fourth generation. 
However there are many problems with this proposal, the most important
being the lack of a
fundamental interaction leading to the condensation of just the fourth
generation and a mechanism to generate the masses of the non-condensing
fermions. In Ref.~\cite{bd} this idea
was pursued in the framework of a 5D model in
anti--de Sitter (AdS)~\cite{Randall:1999ee}. There it was shown that it provides an ultraviolet
(UV) completion of the original model up to a scale of order $M_{Pl}$.
The Kaluza-Klein (KK) partners of the gluon naturally induce the
strong interactions responsible for the condensation of the zero-modes
of the fourth generation. The inevitable presence of bulk four-fermion
operators also contribute to the strong interactions of the fourth
generation and lead to fermion masses after the condensation. The
model naturally explains the fermion mass hierarchy and can be
extended, for instance, by allowing more than one fourth-generation
fermion to condense,  or by modifying the fermion 
embedding into the larger 5D gauge symmetry. Regardless of the
details of the particular realization chosen, there are some generic
implications of this class of models that remain common to all of
them. Among these features are a rather heavy fourth generation, with
masses that can be in the range  $(300-700)~$GeV depending on which quarks
condense, and a heavy Higgs which in the case that only one quark 
condenses has a mass of approximately $m_h\sim (700-900)~$ GeV. 
In the specific realization of this class of models proposed in
Ref.~\cite{bd}, only the up-type quark condenses leading
to a one-Higgs-doublet model.   
We will take it as a benchmark of a
large class of theories where EWSB is triggered in this way, 
in the hope that it can be used to study generic properties and 
signatures that rely only on the existence of a strongly coupled fourth
generation.

The bounds from electroweak precision measurements on a fourth generation have been reanalyzed
recently in Ref.~\cite{Kribs:2007nz}, and are not as tight as they once were. 
More important, in the class of models we are interested in here 
the loop contribution from
the fourth generation to the $S$ parameter (from which the bounds
mainly come) is neither the leading one nor well defined. This can be
seen by the fact that AdS$_5$ models with Planck-brane localized light
fermions, have a large {\em tree-level} $S$
parameter~\cite{sinrs}. Furthermore, it has been recently pointed out 
that in these models the one-loop contributions to $S$ contain
logarithmic divergences and therefore $S$ must be
renormalized~\cite{Burdman:2008gm}. This is also seen in deconstructed
versions of Higgsless models~\cite{treesite}. The renormalization procedure
would then affect all one-loop corrections in such a way that 
it is not correct to use them to put strict bounds on the
contributing states. We then conclude that the presence of a 
fourth generation does not make the $S$-parameter problem of these
models any worse than it is with only three generations.

In this paper we study experimental signatures at the LHC
of the quark sector of a strongly coupled fourth-generation. 
The defining aspect of  these theories is  the presence of a new interaction 
coupling strongly 
to the heavy fourth generation. In particular, the fourth-generation
quarks couple strongly to a 
color-octet vector current, which is responsible for the condensation
of at least one of them. In the bulk AdS$_5$ model of~\cite{bd}, this
current corresponds to the KK excitations of the gluon,  $G^{(n)}$. 
An unmistakable signal for the presence of this strong interaction
would be to observe the production of fourth-generation quarks
in channels involving the KK gluons such as\footnote{There will also be pair 
production via the
  interactions with color-singlet weak-currents corresponding to the
  KK excitations of the electroweak gauge bosons, but these will be suppressed compared with the
  strong production. Thus we will leave this channel for a future
  study and ignore it in the present work. In any case this is a
  conservative assumption.} 
\be
q\bar q\rightarrow G^{(1)}\rightarrow D_4\bar
D_4, U_4\bar U_4
\ee
In the specific scenario we study here, the
$U_4$ is assumed to be the only condensing quark, making it somewhat heavier than 
$D_4$.  
These processes will generate an excess in
the production of the fourth generation when compared with the usual standard model 
(SM) QCD production, that is characteristic of this class of models. This
excess would
provide evidence that the condensation mechanism is associated to 
EWSB.  In particular, since $D_4$ does not
condense, we expect it to be somewhat lighter than $U_4$, and as a
consequence it will be easier to produce $D_4$ pairs than $U_4$ pairs.
Furthermore, when produced the $U_4$ would almost always decay to $D_4$ 
through the charged current. 
We will study the pair production of $D_4$'s both via QCD and
the KK gluons in the model mentioned above. 

In order to define the final state signal, we consider the fact that 
in these models the fourth generation has typically  larger mixing with the 
third generation than with the lighter first two. 
This implies that $D_4$ will mostly decay to $W^-t$. Thus, the pair production
of $D_4$ will lead to events with two $W^+$'s, two $W^-$'s and two
$b$ jets. Final states with only one charged lepton or with two opposite-sign
leptons would be hard to observe at the LHC, above
the large $t\bar t+jets$ background. 
Instead, we study the process with two
same-sign leptons in the final state, which has a much smaller
$t\bar t$ SM background~\cite{Contino:2008hi}. 
We make a preliminary study for the observation of the 
down-type fourth-generation quark in this channel
and find that a 
5$\sigma$ significance requires an accumulated luminosity of about   
$L_{min}\lae{\cal O}(1)~fb^{-1}$ for $m_{D_4}=(300-600)$~GeV. We also study  
the possibility of measuring the excess in $D_4$ from the contribution of 
an s-channel KK gluon above the standard QCD production.
This turns out to be quite difficult since the width of the KK gluon in fourth-generation 
models is of the order of its mass, making the KK gluon excess over the QCD 
production featureless.
As we will see below, a very large data sample, together with an excellent 
understanding of the QCD production process, will be necessary in order to observe 
this excess with significance. 

Other studies of the production and decay of fourth-generation quarks
at the LHC exist in the literature. For instance, in
Ref.~\cite{Holdom:2007nw} a new technique for the heavy quark mass 
reconstruction is discussed, whereas in Ref.~\cite{hou1}
flavor-violating decays involving the fourth generation are
considered. In some of the previous papers~\cite{cernrep} it is assumed that $D_4$ mixes
primarily with the first two generations, instead of with the third as
we consider here. The consequences of the existence of a fourth generation 
in flavor physics have also received considerable attention~\cite{flavor}.

In the next Section we present an effective model describing the
interactions between fermions and vector currents at low
energies. In Section~\ref{strategy} we describe our strategy to
isolate the signal from the SM backgrounds. We then show our results 
regarding the observation of a heavy fourth generation at the LHC in 
Section~\ref{analysis}.
In Section~\ref{kkgluon} we discuss the potential for the separation of the KK gluon 
signal from the standard QCD fourth-generation production, and finally conclude
in Section~\ref{conclusions}.

\section{Effective model}\label{effmodel}

We consider a four-dimensional (4D) theory containing two sectors: a strongly
coupled field theory (SCFT) sector and a sector of elementary fields corresponding
to the SM gauge bosons and fermions, including the fourth generation
\footnote{Although, as we will see below, heavier fermions will be mostly
composite}. The SCFT sector has a large number of colors $N$, and  is conformal at high 
energies. At the low energy 
$M_{IR}\sim {\cal O}(1)~$TeV, conformal invariance is spontaneously broken generating a mass gap.
leading to a discrete spectrum of particles with
the lightest masses being of order of 1~TeV. We assume that the SCFT  has a global symmetry
SU(3)$_c\times$SU(2)$_L\times$SU(2)$_R\times$U(1)$_X$, that contains
the SM gauge symmetry plus an extra $SU(2)$, introduced to preserve 
a custodial symmetry. Hence the operators and states of this sector
furnish complete multiplets of the large global group. This implies
that the SCFT sector has several conserved currents
transforming as color octets or isospin triplets. 
The SM vectors gauge the SU(3)$_c\times$SU(2)$_L\times$U(1)$_Y$
subgroup, with $Y=T^{3R}+X$, and couple to the SCFT through its
conserved currents. We will assume that the SM fermions couple
linearly to the SCFT through fermionic operators ${\cal O}_\psi$
\begin{equation}\label{psiO}
{\cal L}=\lambda \bar{\psi} {\cal O}_\psi + h.c. \ ,
\end{equation}
which are in complete representations of the global symmetry. The SM
fields can be embedded into the corresponding representation of the
larger global symmetry by considering additional non-dynamical
components.
The low energy behavior 
of the running coupling $\lambda$ in Eq.~(\ref{psiO}) is determined by the anomalous 
dimension of the operator ${\cal O}_\psi$, $\gamma=\rm{dim}[{\cal O}_\psi]-5/2$, with 
$\rm{dim}[{\cal O}_\psi]$ the conformal dimension of 
${\cal O}_\psi$~\cite{Contino:2004vy,Agashe:2004rs}. 
For $\gamma>0$ the 
coupling between the elementary fermion and the SCFT is irrelevant, thus at energies 
below $M_{IR}$ we have $\lambda\sim(M_{IR}/\Lambda)^\gamma$, resulting in a small mixing. 
For $\gamma<0$ the coupling is relevant and flows to a fixed-point, resulting in a 
large mixing of the elementary fermion with the SCFT.

The general setup described above provides the tools for a scenario
where the electroweak symmetry is broken by the condensation of the fourth
generation~\cite{bd}. There are at least two sources that can induce the four
fermion interaction needed for the condensation. First we consider
composite operators in the SCFT coupling four fermionic resonances 
(for example $\bar{\cal O}_L {\cal O}_R \bar{\cal O}_R {\cal O}_L$).
These operators induce, through the interactions of Eq.~(\ref{psiO}),
four fermion operators for the SM fermions. Second, there are
operators in the SCFT that couple the fermionic resonances to the
vector resonances created by the conserved global currents. Through
Eq.~(\ref{psiO}), these operators generate interactions between four
elementary fermions by the exchange of vector resonances. The strength
of these interactions is governed by the anomalous dimensions
$\gamma$, that can correctly select the fermions with large
interactions. Some of these fermions may condense breaking the electroweak 
symmetry and generating a dynamical Higgs at low
energies. For simplicity, we will consider a
scenario where just the up quark of the fourth generation, $U_4$,
condenses. Fermion masses result from the four fermion
interactions by considering two operators corresponding to the
condensing fermions. Therefore,
heavy fermions, such as the top quark and the fourth generation, have
large mixings with the SCFT states and thus they will be mostly
composite, whereas the light fermions will be mostly elementary.

Inspired by the AdS/CFT correspondence, Ref.~\cite{bd} 
proposed a weakly coupled 5D realization of the 4D theory described
above. It makes use of a Randall-Sundrum
spacetime~\cite{Randall:1999ee} consisting of a slice of AdS$_5$
with a curvature $k\sim M_{Pl}$. The extra dimension $z$ is compact
with boundaries in conformal coordinates given by $z_0=1/k$ (called
UV boundary) and $z_1=1/M_{IR}$ (called IR boundary). The theory is
defined on the segment $z_0\leq z\leq z_1$. There is a 5D gauge
symmetry $SU(3)_c\times SU(2)_L\times SU(2)_R\times U(1)_X$ that is
broken down to the SM group by boundary conditions on the UV boundary. There
are four generations of 5D fermions that fulfill complete
representations of the bulk gauge symmetry. Boundary
conditions are imposed such that in the UV boundary just the SM fermions and a
standard fourth generation are dynamical. The IR boundary conditions
lead to a set of massless zero modes corresponding to four generations
of SM fermions. The 5D fundamental parameter that determines
the degree of compositeness of the zero modes is the 5D fermion mass
$m_\psi=c_\psi k$. Following the holographic procedure of
Ref.~\cite{Contino:2004vy}, the anomalous dimension associated to a
left-handed fermion is  
$\gamma=|c+1/2|-1$, resulting in  $q_L$ being mostly  fundamental for $c_q\geq 1/2$
and mostly composite for $c_q\leq 1/2$ (and similarly for the
right-handed fields, by making the replacement $c_q\rightarrow -c_{u,d}$)~\footnote{In
  the 5D picture, a composite fermion is localized in the IR boundary
  at the TeV, whereas a fundamental fermion is localized in the
  UV boundary at the scale $M_{Pl}$.}. Therefore, $c_\psi$ is the
fundamental parameter that sets the strength of the couplings between
the fermion $\psi$ and the heavy states. The holographic prescription
allows us to identify the elementary fields with the fields supported
in the UV boundary, and the SCFT dynamics with the bulk and IR degrees
of freedom. In this way, the KK modes are the resonances of the SCFT.

In Ref.~\cite{bd} a bulk four-fermion
interaction was considered, and its coefficient estimated by naive dimensional analysis
(NDA). This operator leads
to the four-fermion interaction in the SCFT sector mentioned above.
On the other hand, the bulk gauge symmetry gives rise to 4D conserved
currents, associated with the KK modes of the 5D gauge fields, that couple 
to the fermionic modes.  Relying on the results of
Ref.~\cite{bd}, we estimate that the four fermion
interaction induced by the color-octet current mediation is of the same order
(although numerically somewhat larger) as the one induced by the 5D
four-fermion operator (see Eq.~(\ref{ceff}) below). The zero modes of
the fourth generation condense if they are strongly localized towards
the IR boundary, meaning that they are mostly composite states of the
SCFT.
 
In what follows we consider the effective Lagrangian describing
the relevant interactions for the processes we want to study. 
In the language of the weakly coupled 5D theory 
the relevant degrees of freedom correspond to the fermionic zero modes,
the SM gauge fields and the first KK-vector resonances. At low
energies the four fermion interactions and the interactions with the
KK vectors are described by:
\begin{equation}\label{l1}
{\cal L}=M_1^2Tr[G_\mu^{(1)}G^{(1)\mu}]+\sum_{\psi=q_l,u_R,d_R}g_{s\psi^a}\bar \psi^a\gslashkk \psi^a
+\sum_{\psi\psi'=u,d}C_{abcd}(\bar q_L^a \psi_R^b) (\bar \psi_R^{'c} q_L^d)+h.c. \, ,
\end{equation}
\noindent 
where $G_\mu^{(1)}$ is the first KK gluon, $a=1,\dots,4$
numbers fermion generations, and we have neglected the momentum of the
KK vectors compared with their mass. The coupling $g_{s\psi}$
corresponds to the one
between the first KK excitation of the gluon and the zero mode of the fermion $\psi$,
and it depends on the degree of compositeness of the fermion, or its
localization in the extra dimension in the 5D picture. For a 5D
model $g_{s\psi}$ varies between $\simeq 8.4\,g_s$, corresponding
to a composite (or TeV-localized) fermion,  and $\simeq -0.2\,g_s$, corresponding to a fundamental
(or Planck localized) fermion~\cite{Gherghetta:2000qt}. The mass of
the first KK gluon $M_1$
depends on the size of the extra dimension $1/M_{IR}$, and is
approximately given
by $M_1\simeq 2.4 M_{IR}$. The Higgs mass and the mass of the
condensing quark $U_4$ depend on $M_{IR}$. For instance, for
$M_{IR}\simeq 1~$TeV one has~\cite{bd} $m_h\simeq 900~$GeV and
$m_{U_4}\simeq 700$ GeV.  
There are also similar interactions
involving the KK modes of the electroweak gauge bosons that we have not written explicitly.
Their structure is the same as the one for the KK gluon, with the
couplings normalized with respect to the electroweak couplings $g$ and
$g'$, instead of the QCD coupling $g_s$.

The last term of Eq.~(\ref{l1}) contains the four-fermion interaction
that generates the mass terms of the non-condensing fermions after the
condensation of $U_4$. The coefficient $C_{abcd}$ is a dimensionfull parameter that
depends on all the 5D masses $c_\psi$ of the fermions involved in the
interaction. This coupling is exponentially suppressed if at least one
of the fermions is Plank-brane localized. This mechanism allows us to obtain the
top-bottom hierarchy with fundamental parameters of the same order.
The exact value of  $C_{abcd}$ depends on the embedding of the
SM fermions into the larger bulk gauge group, but up to numbers of
${\cal O}(1)$, we expect it to be independent of the details of the
model. In any case and to fix things, we assume that the 5D fermions transform as: $q\in({\bf
  2},{\bf 2})_{2/3}+({\bf 2},{\bf 2})_{-1/3}$, $u\in({\bf 1},{\bf 1})_{2/3}$ and $d\in({\bf
  1},{\bf 1})_{-1/3}$ of SU(2)$_L\times$SU(2)$_R\times$U(1)$_X$, as
dictated by the constraints on the $Zb_L\bar b_L$
couplings~\cite{Agashe:2006at} (the case with $u,d\in({\bf 3},{\bf
  1})_{2/3}+({\bf 1},{\bf 3})_{2/3}$ is very similar and we do not
expect large corrections in the process we are working on). 
The four-fermion coupling is then given by~\cite{bd}
\begin{eqnarray}\label{ceff}
C_{abcd}=& &\,C_{abcd}^{5D}\,\,\frac{k^3}{M_{Pl}^3 M_{IR}^2}\,\,
\frac{1-x^{4- c_L^a + c_R^b + c_R^c - c_L^d}}{4 - c_L^a + c_R^b + c_R^c - c_L^d} 
\nonumber \\
&\times& \left[\frac{(1-2c_L^a)(1+2c_R^b) (1+2c_R^c)(1-2c_L^d)}
{(1-x^{1-2c_L^a})(1-x^{1+2c_R^b})(1-x^{1+2c_R^c})(1-x^{1-2c_L^d})}\right]^{1/2}
\, ,
\end{eqnarray}
\noindent where $x=M_{IR}/k$ and  $C_{abcd}^{5D}$ is a dimensionless coefficient
measuring the strength of the four-fermion interaction in the 5D
theory. We can estimate its size within NDA, as being
\be
C_{abcd}^{5D} \simeq O(1)\,\frac{36\pi^3}{N}~,
\label{nda}
\ee
where $N$ is the number of fermions running inside a 5D loop
diagram contributing to the four-fermion interaction. 
For our specific choice for the embedding of the quarks into
the larger 5D gauge symmetry, $N=400$.

Since the Higgs comes from the condensation of the mostly-composite $U_4$, 
the size of the Yukawa couplings of the non-condensing
fermions would also depend on their degree of compositeness or
localization in the 5D. This means that
the Yukawas are also controlled by $c_\psi$. To obtain a heavy but
non-condensing $D_4$, this fermion must have a rather large degree of
compositeness, but with an upper bound so as to forbid its condensation.
This roughly means $|c_{D_4}|<1/2$. The mass of $D_4$ depends on
the value of $c_{D_4}$, but we estimate it to be in the range
$(200-600)~$GeV. In order to avoid direct bounds~\cite{pdg,pqmarc} we conservatively
consider here $m_{D_4}\geq 300$~GeV.

Another necessary consideration is the mixing of the fourth-generation
with the other three.
The analysis of
Ref.~\cite{Kribs:2007nz} shows that there are strong
constraints in the mixings between the light fermions and a fourth
generation. On the other hand, the $95\%$ C.L. lower bound
$V_{tb}>0.68$, obtained from the observation of single top
production~\cite{Abazov:2006gd}, still allows for a large mixing between
the third and the fourth-generation quarks. For the purpose of
our work, we need only assume that this mixing is much larger than the ones with
the two lighter generations, resulting in the dominant decay mode of
$D_4$ being $D_4\to W^-t$.

The interactions of Eq.~(\ref{l1}) lead to the decay of the KK gauge
bosons and determine their widths and branching ratios. The width of
the first KK gluon is mostly determined by its couplings to the
fourth-generation quarks, and to a lesser extent to the top quark,
since its couplings to light quarks are much smaller. We have scanned
over the parameter space of the 5D model, with the following
constraints: $U_4$ has a supercritical four-fermion effective
interaction and condenses, $D_4$ is heavier than $300~$GeV but does
not condense, the top and bottom have their physical masses. We have
considered 5D parameters $C_{abcd}^{5D}$ not larger than $3$ times the
NDA estimate of Eq.~(\ref{nda}) in order to avoid unnaturally large
numbers in the fundamental theory. The width of the first KK gluon,
$\Gamma_1$, is larger the more composite (or TeV-localized) are $U_4$
and $D_4$. On the other hand, $\Gamma_1$ is smaller the closer is
the $U_4$ coupling from being critical, and the lighter is
$D_4$. However, in most cases the width is typically $\Gamma_1\sim
M_1\simeq (2.4-3)$~TeV. We can consider the lower limit to be
$\Gamma_1\simeq 0.37 M_1$, obtained for $C_{4444}^{5D}$ three times
its NDA estimate, tuning the effective 4D four-fermion interaction to
be in the critical limit and taking $D_{4R}$ to be almost
fundamental. However, we stress that this is only a very small region
of the parameter space, and the natural size of the KK-gluon width is
$\Gamma_1\sim M_1$. This is due to its large couplings to the fourth
generation, as well as to the top. This result has important
consequences for the phenomenology, since it precludes the existence
of a resonant peak associated with the color-octet current responsible
for $U_4$ condensation, and ultimately for electroweak symmetry
breaking.

The pair production of the fourth-generation quarks proceeds mostly through
QCD and KK-gluon mediation~\footnote{There is also a contribution
mediated by the four fermion interaction. However it is
exponentially suppressed, see Eq.~(\ref{ceff}).}. In what follows, we
discuss the strategy to observe the signal, as well as the possibility
of separating the KK gluon contribution from the standard QCD
production mechanism.

\section{Signal and backgrounds}
\label{strategy}

\begin{figure}
\begin{center}
\begin{picture}(250,110)
	\ArrowLine(-5,70)(35,50)
	\ArrowLine(35,50)(-5,30)
	\Vertex(35,50){3}
	\Gluon(35,50)(80,50){3}{6}
	\Vertex(80,50){3}
	\ArrowLine(80,50)(200,100)
	\ArrowLine(200,0)(80,50)
	\ArrowLine(97,57)(102,59)
	\ArrowLine(102,41)(97,43)
	\ArrowLine(180,92)(185,94)
	\ArrowLine(185,6)(180,8)
	\Vertex(114,63){3}
	\Photon(114,63)(170,63){3}{6}
	\Vertex(114,37){3}
	\Photon(114,37)(170,37){3}{6}
	\Vertex(170,63){3}
	\ArrowLine(170,63)(210,73)
	\ArrowLine(210,58)(170,63)
	\Vertex(165,83){3}
	\Photon(165,83)(220,83){3}{6}
	\Vertex(165,17){3}
	\Photon(165,17)(220,17){3}{6}
	\Vertex(170,37){3}
	\ArrowLine(210,27)(170,37)
	\ArrowLine(170,37)(210,42)
	\Vertex(220,83){3}
	\ArrowLine(220,83)(260,93)
	\ArrowLine(260,73)(220,83)
	\Vertex(220,17){3}
	\ArrowLine(260,7)(220,17)
	\ArrowLine(220,17)(260,27)
	\Text(55,63)[c]{$g,G^{(1)}$}
	\Text(-10,70)[c]{$q$}
	\Text(-10,30)[c]{$\bar q$}
	\Text(100,70)[c]{$D_4$}
	\Text(100,30)[c]{$\bar D_4$}
	\Text(155,72)[c]{$W^-$}
	\Text(155,47)[c]{$W^+$}
	\Text(205,92)[c]{$W^+$}
	\Text(209,8)[c]{$W^-$}
	\Text(219,73)[c]{$q$}
	\Text(219,58)[c]{$\bar q$}
	\Text(219,42)[c]{$l^{'+}$}
	\Text(219,27)[c]{$\nu$}
	\Text(145,87)[c]{$t$}
	\Text(145,13)[c]{$\bar t$}
	\Text(188,105)[c]{$b$}
	\Text(188,-5)[c]{$\bar b$}
	\Text(269,93)[c]{$l^+$}
	\Text(269,73)[c]{$\nu$}
	\Text(269,27)[c]{$q'$}
	\Text(269,7)[c]{$\bar q'$}
\end{picture}
\caption{Feynman diagram corresponding to the pair production of $D_4$ and
  decay to a final state with two same-sign leptons. 
}
\label{figfeynman}
\end{center}
\end{figure}
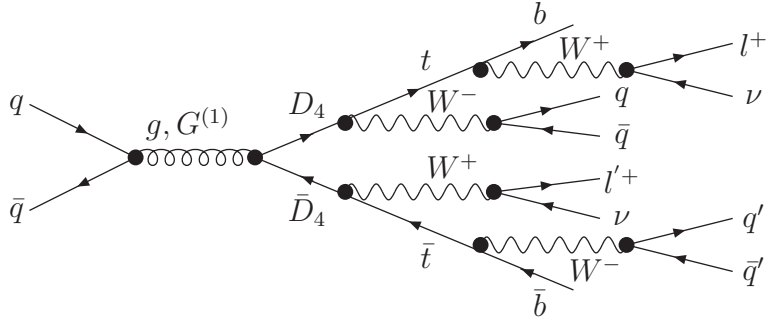

The pair production of $D_4$ can proceed through the standard QCD  
contributions to heavy quark productions, as well as through the
s-channel contribution of the first KK excitation of the gluon. 
With the assumption of the mode $D_4\to tW^-$ dominating the decay,  
the final state contains multiple $W$'s
\begin{equation}
  pp \to D_4 \bar{D}_4 \to W^-t\;\; W^+\bar{t} 
                       \to  W^+ W^-b\;\; W^+ W^-\bar{b}  
\label{finalstate}
\end{equation}
rendering the full reconstruction of the final state difficult.
We then choose to focus on inclusive states exhibiting
charged leptons ($e^\pm$ and $\mu^\pm$) from $W$ decays and jets; see
Figure~\ref{figfeynman}.  More specifically we study the production
of two--, three--, and four--leptons accompanied by at least two hard
jets. In the case of dilepton production the standard model $t
\bar{t}$ production is a formidable background. As a consequence, we
restrict our analysis to same-sign dileptons allowing them to be of
different flavor.  The same-sign dilepton inclusive channel turns out
to be one of the best modes to study the $D_4$ pair
production, being a powerful tool to falsify our model in case that
no events are observed. 
However, an observation of this  particular signal should be supplemented by 
others so as to confirm the $D_4$ hypothesis. 
The processes with three and four leptons 
should also be observed if there is a heavy fourth
generation.  As we will see below, for the four-lepton processes, the smaller signal is
compensated by a negligible background. 
In conclusion, this set of signals offers a very good
test for the model which should be eventually completed with the
full reconstruction of the $D_4$ mass peak in its decay into jets.

The production of $D_4$ pairs at the LHC takes place in $q \bar{q}$ fusion, via
$g$ and $G^{(1)}$ exchange, as well as, QCD gluon--gluon fusion.  
\begin{eqnarray}
  &q\bar q \rightarrow (G^{(1)}\; , \; g) &\rightarrow D_4 \bar D_4 
\label{kkprod}
\\
 &g g   &\rightarrow D_4 \bar D_4 
\label{qcdprod}
\end{eqnarray}
In order to obtain the final states that can be reached from the above
processes, {\em e.g.} Eq.~(\ref{finalstate}), we work in the narrow width
approximation for the $D_4$'s, as well as for the top quarks and $W$'s coming
from its decay. Notwithstanding, we preserve the spin correlations in the
production and decay chains of the new fourth generation quarks.

Same-sign dileptons are produced when two same-charge $W$'s decay
leptonically while the two other $W$'s decay hadronically.  In general
these events present a large number of jets and significant missing
transverse momentum due to escaping neutrinos. In the dilepton search
we require at least two hard jets, however, we did not impose any cut on
the observed transverse momentum or try to explore the jet multiplicity of the
events since the same sign dilepton
signal is extremely clean. The main SM backgrounds are:
\begin{itemize}

\item QCD production of $t \bar{t}$ and its decay into $W^+W^- b \bar{b}$.  In
  this channel one of the same-sign leptons originates from a $W$ while the
  other lepton comes from the semi-leptonic decay of the $b$. Although 
there is a small probability to obtain an isolated lepton from the $b$ decay,
this is compensated by the large production cross section.

\item $W^\pm W^\pm jj$ production followed by the leptonic decay of the same
  charge $W$'s. Here $j$ denotes a jet.

\item $W^\pm W^\pm jjj$: although this process is higher order in QCD
  with respect of the previous one, this can be compensated if the extra
  jet is not very hard\footnote{We require that the additional jet in
    the event passes the acceptance and isolation cuts given in
    Eq.~(\ref{acciso}).}.

\item $W^\pm Z jj$ where the weak gauge bosons decay leptonically and 
the differently charged lepton escapes undetected.

\item  $W^\pm t\bar t$ where one of the top quarks decay semi-leptonically
while the other decay into jets.

\item We also considered the following possible sources of same sign
  dileptons: $W^\pm W^\pm W^\mp $ with and without an extra jet and $W^+ W^- t
  \bar{t}$ production

\end{itemize}
We used MadEvent~\cite{Alwall:2007st} to generate the signal and above
backgrounds at the parton level, except for the $t \bar{t}$ production
that was studied using PYTHIA version
6.409~\cite{Sjostrand:2000wi,Sjostrand:1993yb} in order to better take
into account the semi-leptonic decay of the $b$ quark.


In the trilepton signal only one $W$ in Eq.~(\ref{finalstate}) decays
hadronically while the others decay leptonically. These events present a
smaller jet activity than the dilepton signal, although we still require 
the presence of two hard jets. The main SM backgrounds for the trilepton
channel are: 
\begin{itemize}

\item diboson electroweak gauge boson production, {\em i.e.} $ZZjj$ and
  $W^\pm Zjj$,  where the $W$'s and $Z$'s decay leptonically;

\item $W^\pm W^\pm W^\mp jj$ that receives a contribution from
the intermediate state $t \bar{t} W^\pm$ when the jets are $b$ jets.


\end{itemize}

Finally, the cleanest state that can be obtained from Eq.~(\ref{finalstate}) is
when all $W$'s decay into leptons. Although only a small fraction of the
signal ends up in this state, this is compensated by an
extremely low background. We looked for this topology requiring four
leptons and two jets in the central region of the detector. In our
study we took into account the main SM backgrounds: $ZZjj$,
$W^+W^-Zjj$, and $W^+W^+W^-W^-jj$ productions.

We present our results for three representative points of the
parameter space given by three $D_4$ masses: 300, 450, and 600 GeV. Nevertheless,
a choice for the $D_4$ mass does not completely fix the parameter space
because there is still some freedom in obtaining the couplings and width of
the first Kaluza--Klein excitation of the gluon~\cite{bd}.  We further assume
a heavy Higgs with $m_h\sim 900$ GeV, which is consistent with the existence of
a condensing heavy fourth generation. In Table~\ref{tparam} we show the
first KK--gluon width and couplings used in our simulations which were
computed using the 5D condensation model~\cite{bd}.  The large values of
the couplings seen in this table correspond to the fourth generation being
almost completely composite. Although we quote here the results for only three benchmark
points in the parameter space of the model, we scanned over a large region of
the parameter space and we checked that our results are general enough not
depending on our specific choices.

\begin{table}[h]
\begin{center}
  \begin{tabular}{| c | c | c | c | c | c | c |} 
    \toprule
    $m_{D_4}$[GeV] & $g_{sQ^4_L}$ & $g_{sU_{4R}}$ & $g_{sD_{4R}}$ & $g_{sq^3_L}$
    & $g_{st_R}$ & $\Gamma_1/M_1$\\ 
    \midrule
    300   & $4.4 g_s$ & $4.4 g_s$ & $1.1 g_s$ & $0.5 g_s$ & $3.3 g_s$ & 0.68 \\
    \midrule
    450   & $3.7 g_s$ & $4.4 g_s$ & $2.8 g_s$ & $0.6 g_s$ & $3.9 g_s$ & 0.68 \\
    \midrule
    600   & $5.4 g_s$ & $5.4 g_s$ & $3.1 g_s$ & $1.0 g_s$ & $1.7 g_s$ & 0.98 \\
    \bottomrule
  \end{tabular}
\end{center}
\caption{Benchmark points in the parameter space of the effective theory. 
  $g_{s\psi}$ is the coupling between the first KK gluon and the fermion 
  $\psi$, and depends on the degree of compositeness of the fermion.}
\label{tparam}
\end{table}


\section{Analysis}\label{analysis}

In this Section we study the two, three, and four lepton signals
from $D_4$ pair production and their respective SM backgrounds, in order to
assess the LHC potential for the discovery of this new heavy quark. We
quantify the LHC reach by quoting the minimum required luminosity for a given signal
to be observed.

\subsection{Signal with two same-sign leptons}\label{analysis2l}


The $D_4$ pair production leads to same-sign dileptons when two equally
charged $W$'s decay leptonically. Although it does not allow to reconstruct the
$D_4$ mass peak, this topology presents only a modest SM
background. Therefore, this final state should provide a first hint of the
existence of the fourth generation. We start by applying the following
acceptance and isolation cuts:
\begin{equation}
\begin{array}{lll}
  p_T^\ell > 10 \hbox{ GeV} \;\;,\;\; & |\eta_\ell|<2.5 \;\;,\;\; & 
  \\
  p_T^j >20  \hbox{ GeV} \;\;,\;\; & |\eta_j |<3 \;\;,\;\; &
  \\
  \Delta R_{\ell\ell}\geq 0.7 \;\;,\;\; & \Delta R_{\ell j}\geq 0.7 \;\;,\;\; & 
      \Delta R_{jj}\geq 0.7 \;\;,\;\;
\end{array}
\label{acciso}
\end{equation}
where $\ell$ ($j$) are the two hardest charged leptons (jets). In
Table \ref{tcross} we
present  the cross sections after the above cuts for the
signal and SM backgrounds. For the signal we quote the results
with and without the inclusion of the KK gluon contribution in order to
estimate its impact on  $D_4$ pair production.
\begin{table}
\begin{center}
  \begin{tabular}{| l | c | c |  c |} 
\toprule
   process/cuts & (\ref{acciso}) & (\ref{acciso}) and (\ref{cutjet}) 
&(\ref{acciso}), (\ref{cutjet}), and (\ref{cutlep}) 

\\ 
\midrule
  signal: $m_{D_4}=300$ GeV & 1388 & 412    &  87.0 \\
\midrule
  QCD: $m_{D_4}=300$ GeV    & 1360 & 402    & 83.4 \\
\midrule
  signal: $m_{D_4}=450$ GeV & 222  & 164    & 54.2 \\
\midrule
  QCD: $m_{D_4}=450$ GeV    & 204  & 150    & 48.8 \\
\midrule
  signal: $m_{D_4}=600$ GeV & 48   & 44     & 17.8 \\
\midrule
  QCD: $m_{D_4}=600$ GeV    & 42   & 38     & 15.5 \\
\midrule
  $t\bar t$                 & 2060 & 452    & 1.2 \\
\midrule
  $W^+W^+jj$                & 8.2  & 4.0    & 1.0 \\
\midrule
  $W^-W^-jj$                & 3.8  & 1.8    & 0.6 \\
\midrule
  $W^+W^+jjj$               & 8.1  & 5.0    & 1.3 \\
\midrule
  $W^-W^-jjj$               & 3.4  & 2.1    & 0.8 \\
\midrule
  $W^+Zjj$                  & 9.4  & 1.2    & 0.3 \\
\midrule
  $W^-Zjj$                  & 4.3  & 0.6    & 0.2 \\
\midrule
  $W^+t\bar t$              & 8.6  & 2.3    & 0.6 \\
\midrule
  $W^-t\bar t$              & 3.5  & 0.9    & 0.2 \\
\midrule
  $W^+W^-t\bar t$           & 0.2  & 0.1    & - \\
\midrule
  $W^+W^+W^-$               & 0.7  & -      & - \\
\midrule
  $W^+W^+W^-j$              & 1.2  & 0.3    & - \\
\bottomrule
  \end{tabular}
\end{center}
\caption{Same sign dilepton signal and SM background cross sections in fb 
  for several choices of cuts.   We present the signal cross section with
  the inclusion of the KK gluon
  contribution,  denoted by signal, and with just the QCD contribution, marked
  as QCD.   The empty boxes correspond to cross sections $\sigma\lesssim{\cal
    O}(10^{-2})$fb.}
\label{tcross} 
\end{table}
We can see from Table \ref{tcross} that the SM background is still quite large after
these minimum cuts, totaling 2.1 pb. Its main contribution comes from 
$t\bar{t}$ production, despite the small probability of obtaining an isolated
lepton from the semi-leptonic decay of a $b$ quark. The strong interaction
production of $t \bar{t}$ pairs account for 98\% of the total SM background.
On the other hand, the signal cross section is appreciable, varying from 50 to
1300 fb depending on the $D_4$ mass.


\begin{figure}[th]
\begin{minipage}{\linewidth}
\epsfig{file=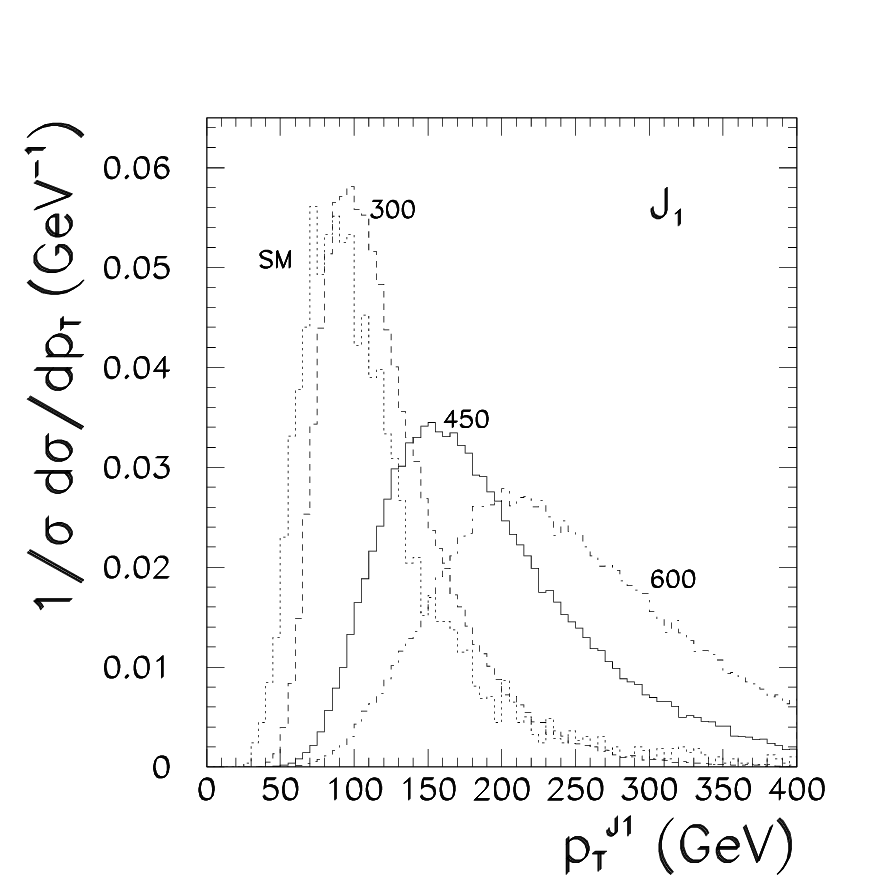,width=0.498\linewidth}
\quad
\epsfig{file=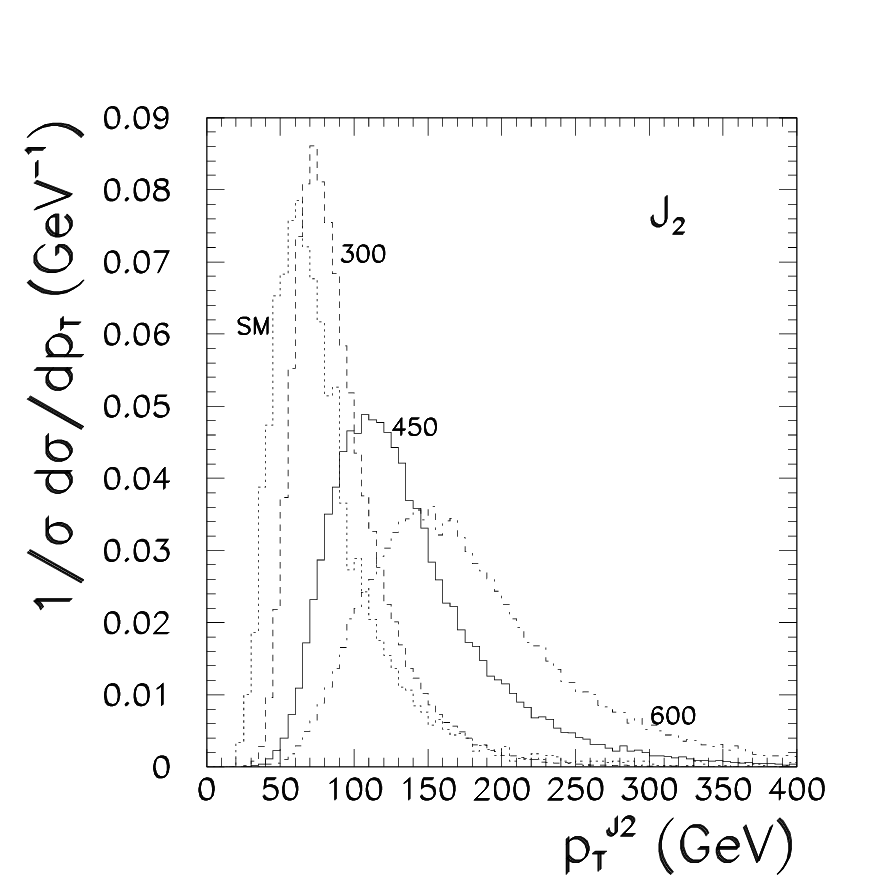,width=0.498\linewidth}  
\end{minipage}
\caption{Normalized transverse momentum distribution of the two hardest jets 
after applying the acceptance and isolation cuts given in Eq.~(\ref{acciso}).
The left (right) panels depicts the $p_T$ distribution
of the hardest (second hardest) jet. 
The dashed, continuous and dotted-dashed lines correspond respectively
to the signal with $m_{D_4}=300,450,600$GeV, the dotted lines
correspond to the sum of the backgrounds, dominated by $t\bar t$.}
\label{fig1}
\end{figure}


In order to suppress the SM backgrounds and enhance the signal we tightened
the transverse momentum cuts. In Figure \ref{fig1} we present  the 
normalized transverse distribution for the two hardest jets in the event.
As it can be seen from this figure, the signal
has a tendency of producing harder jets, especially for larger
masses. Therefore, we further
require that the hardest jets ($j_{1,2}$) satisfy
\begin{equation}
  p_T^{j_{1,2}} > 100 \hbox{ GeV} \; .
\label{cutjet}
\end{equation}
In principle, this cut could be harder for heavier $D_4$'s. However,
here we keep it constant for the sake of simplicity. The effects of the cuts
(\ref{acciso}) and (\ref{cutjet}) in the signal and SM backgrounds are
presented in Table \ref{tcross}.

\begin{figure}[th]
\begin{minipage}{\linewidth}
\epsfig{file=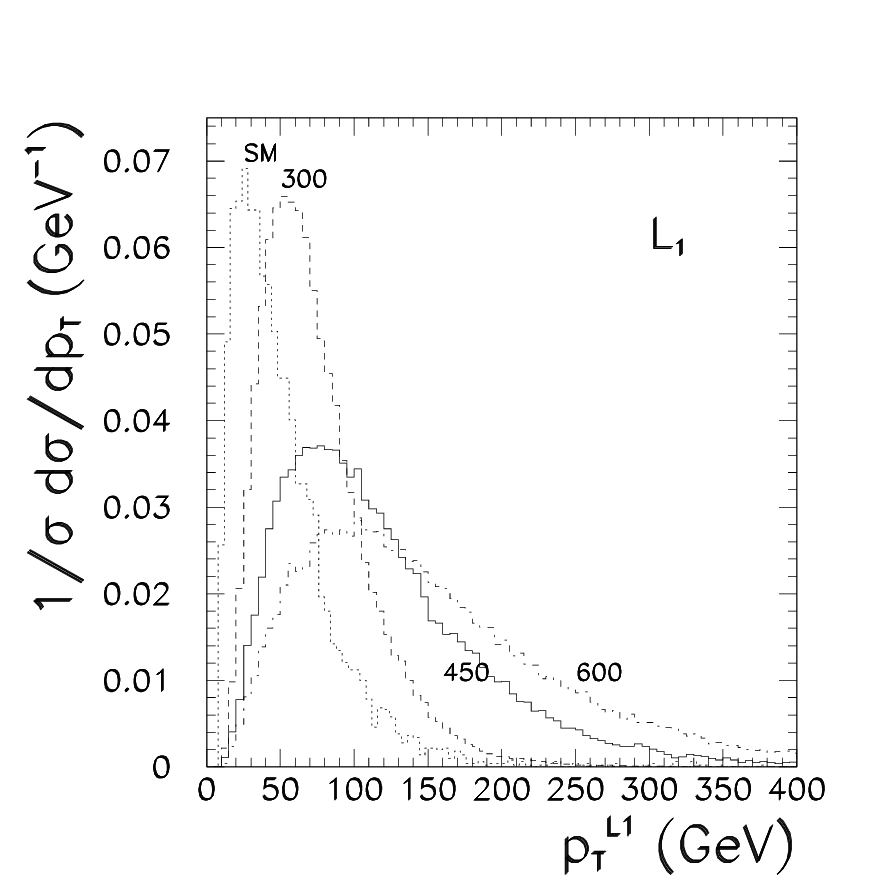,width=0.498\linewidth}
\quad
\epsfig{file=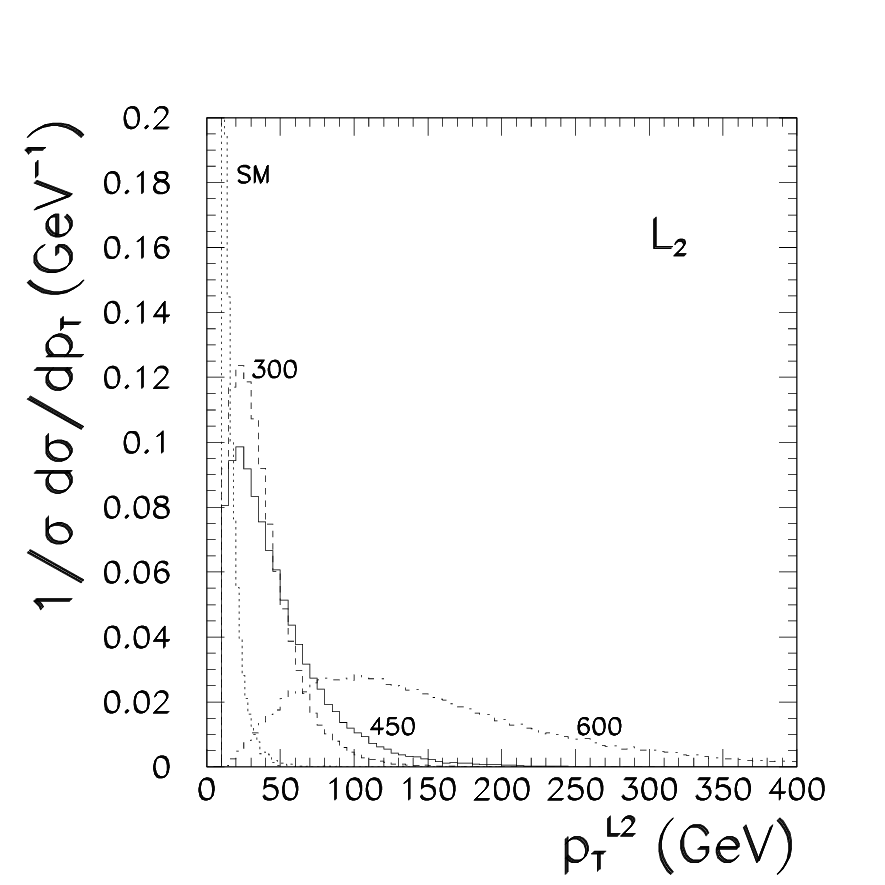,width=0.498\linewidth}  
\end{minipage}
\caption{Same as Fig.~\ref{fig1} but for leptons.}
\label{fig1a}
\end{figure}

Further reduction of the SM backgrounds can be achieved by demanding
harder leptons, since leptons originating from the $b$ semi-leptonic
decay in $t \bar{t}$ production are rather soft. This can be 
seen in Figure \ref{fig1a}, displaying the lepton transverse
momentum spectrum of the signal and backgrounds. In particular, the second hardest
lepton has a very steep spectrum compared to the background, since the
latter is dominated by $t\bar t$. We then  
require that the two same sign leptons satisfy
\begin{equation}
   p_T^{\ell_{1,2}} > 50 \hbox{ GeV} \; .
\label{cutlep}
\end{equation}
This cut has a large impact in the $t \bar{t}$ background, which is
suppressed by a factor $\simeq 370$ as it can be seen from Table
\ref{tcross}. On the other hand, this cut reduces the signal only by
a factor of 3-5.

Table \ref{tsignificance} summarizes the signal and SM background
cross sections after cuts (\ref{acciso}), (\ref{cutjet}) and
(\ref{cutlep}). We  see that the dilepton channel has a large
signal-to-background ratio even for heavier $D_4$'s. We also present
in this table the required integrated luminosity for the signal to
have a statistical significance of $5\sigma$. We conclude that this
signal can be established at the very early stages of the LHC run,
even before collecting 1 fb$^{-1}$. 

\begin{figure}[t]
\begin{center} 
\includegraphics[width=11cm]{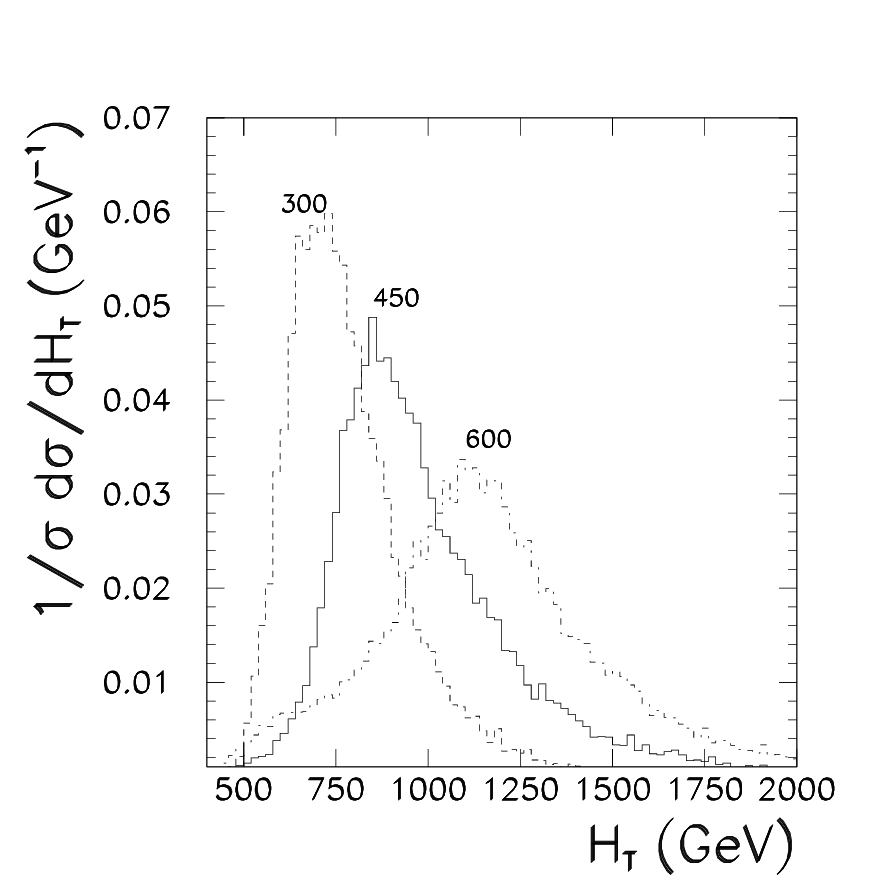}
\caption{Normalized differential cross section as a function of $H_T$
  for the signal with two same-sign leptons. We applied cuts
  (\ref{acciso}), (\ref{cutjet}), and (\ref{cutlep}) and we have
  summed the SM background to the signal. The total area under the curves
  are normalized to one in all the cases.}
\label{fight_3s} 
\end{center}
\end{figure}

Finally, 
the total transverse energy $H_T$ can be used since it follows $m_{D_4}$, even
though it is not possible to reconstruct the $D_4$ mass. The
$H_T$ spectrum is peaked at a value of $H_T$ of order of $H_t\sim 2
m_{D_4}$, as seen in Fig.~\ref{fight_3s}. In order to make a more
quantitative statement, we can fit the mean value\footnote{We have considered
the mean value of $H_T$ instead of the maximum because the former is
much more stable under statistical fluctuations.} 
of  $H_T$, 
$\langle H_t\rangle$,  with $m_{D_4}$  after cuts
(\ref{acciso}), (\ref{cutjet}), and (\ref{cutlep}). We obtained that
\begin{equation}\label{HtmD}
\langle H_T\rangle = a + b \ m_{D_4} \ ,
\end{equation}
with $a = 426$ GeV and $b = 1.23$.  We stress that Eq.~(\ref{HtmD})
gives just a rough estimate, and should not replace a
determination of $m_{D_4}$ obtained from a full $D_4$ mass
reconstruction. 
\begin{table}[h]
\begin{center}
  \begin{tabular}{| c | c | c | c | c|} 
\toprule
 $m_{D_4}$   & $\sigma_{\cal{S}}[{\rm fb}]$ & $\sigma_{\cal{B}}[{\rm fb}]$ & $\cal{S}/\cal{B}$
 & ${\cal L}_{min}$[pb$^{-1}$] \\ 
\midrule
   $300$ GeV & 87.0 & 6.2 & 14. & 44 \\
\midrule
   $450$ GeV & 54.2 & 6.2 & 8.7 & 84 \\
\midrule
   $600$ GeV & 17.8 & 6.2 & 2.9 & 460 \\
\bottomrule
  \end{tabular}
\end{center}
\caption{Same sign dilepton signal and background total cross section, as well as, signal 
to background ratio after cuts (\ref{acciso}),  (\ref{cutjet}), and 
(\ref{cutlep}). ${\cal L}_{min}$ stands for the minimum integrated luminosity
needed to discover the dilepton signal at $5\sigma$ level.}
\label{tsignificance}
\end{table}

\subsection{Trilepton signal}

Given that the  $D_4$ production cross section is large enough, 
confirmation of the fourth-generation origin of the previous signal
could be obtained by the observation of the less-frequent
trilepton events. In this
case, three of the $W$'s present in (\ref{finalstate})
decay leptonically. This final state is clean and presents very
small SM backgrounds. 

We require the presence of three, and only three,
charged leptons and at least two jets satisfying the following
acceptance and isolation cuts
\begin{equation}
\begin{array}{lll}
  p_T^\ell > 10 \hbox{ GeV} \;\;,\;\; & |\eta_\ell|<2.5 \;\;,\;\; & 
  \\
  p_T^j >20  \hbox{ GeV} \;\;,\;\; & |\eta_j |<3 \;\;,\;\; &
  \\
  \Delta R_{\ell\ell}\geq 0.7 \;\;,\;\; & \Delta R_{\ell j}\geq 0.7 \;\;,\;\; & 
      \Delta R_{jj}\geq 0.7 \;\;.
\end{array}
\label{acciso3}
\end{equation}
In Table \ref{tcross3} we collect  the total cross sections for the
signal and SM backgrounds after these cuts. We verified that these
cuts are enough to suppress the background coming from top pair
production followed by the leptonic decay of the $W$'s, with one of the
$b$'s decaying semi-leptonically.
The signal to background ratio is already large after the acceptance
cuts. However, we make further requirements in order to enhance the
signal for large $m_{D_4}$, by demanding that the two hardest jets
satisfy
\begin{equation}
  p_T^{j_1} > 80 \hbox{ GeV} \;\;\;\hbox{ and }\;\;\;
  p_T^{j_2} > 50  \hbox{ GeV} \; .
\label{ptjet3}
\end{equation}
After the initial cuts, the dominant SM background comes from $W^\pm Z jj$
production. This can be efficiently suppressed by imposing
a cut on the invariant mass of the lepton pair of the same flavors and
opposite charges
\begin{equation}
    m_{\ell^+\ell^-} > 100 \hbox{ GeV.}
\label{invmas}
\end{equation}
The effect of these two additional cuts is presented in Table
\ref{tcross3}. As we can see, the SM background becomes negligible
after these cuts while a large fraction of the signal is kept at large
$m_{D_4}$. For small $m_{D_4}$ the signal is suppressed by a factor of
$\simeq 3$. However, the signal cross section is still large enough
even after this reduction.
\begin{table}[h]
\begin{center}
  \begin{tabular}{| c | c | c |} 
\toprule
  process/ cuts & (\ref{acciso3}) & (\ref{acciso3}), (\ref{ptjet3}), 
       and (\ref{invmas}) \\ 
\midrule
  signal: $m_{D_4}=300$ GeV & 612 & 210 \\
\midrule
  QCD: $m_{D_4}=300$ GeV & 604    & 200 \\
\midrule
  signal: $m_{D_4}=450$ GeV & 100 & 61 \\
\midrule
  QCD: $m_{D_4}=450$ GeV & 93.7     & 56.9 \\
\midrule
  signal: $m_{D_4}=600$ GeV & 21.3  & 15.4 \\
\midrule
  QCD: $m_{D_4}=600$ GeV & 1.5     & 13.3\\
\midrule
  $W^\pm Zjj$ & 79.4 & - \\ 
\midrule
  $W^\pm W^\pm W^\mp jj$ & 3.9 & 1.0 \\
\midrule
  $ZZjj$ & 1.3 & - \\
\bottomrule
  \end{tabular}
\end{center}
\caption{
  Trilepton signal and SM background cross sections in fb after cuts.
  We present the signal cross section with the inclusion of the KK gluon
  contribution,  denoted by signal, and with just the QCD contribution, marked
  as QCD.   The empty boxes correspond to cross sections
  $\lesssim{\cal O}(10^{-2})$ fb.
}
\label{tcross3}
\end{table}

Since the trilepton signal is essentially background free after cuts,
we required 5 events to determine the integrated luminosity needed to
establish the signal at the LHC. As shown in Table
\ref{tsignificance3}, the discovery of the trilepton signal requires
approximately the same integrated luminosity as the one needed to establish the
same-sign dilepton signal.  This fact can be used to further tests of
the model.

\begin{table}[h]
\begin{center}
  \begin{tabular}{| c | c | c | c |} 
\toprule
 $m_{D_4}$   & $\sigma_{\cal{S}}[{\rm fb}]$ & $\sigma_{\cal{B}}[{\rm fb}]$ 
& ${\cal L}_{min}$[pb$^{-1}$] \\ 
\midrule
  300 GeV & 210 & 1 &  24 \\
\midrule
  450 GeV & 61.0 & 1 &  82 \\
\midrule
  600 GeV & 15.4 & 1 &  325 \\
\bottomrule
  \end{tabular}
\end{center}
\caption{
  Trilepton signal and background total cross sections 
  after cuts (\ref{acciso3}), (\ref{ptjet3}) and 
  (\ref{invmas}). ${\cal L}_{min}$ stands for the minimum integrated 
  luminosity needed to discover the trilepton signal with the production of
  5 events in the absence of SM backgrounds.
}
\label{tsignificance3}
\end{table}

\subsection{Signal with four leptons}

Finally, $D_4$ pair production also leads to final states with four
isolated leptons and two jets. Although $D_4$ production leads to this
final state only in 0.2\% of the events, the final state is extremely
clean and with a low background. We select these events by applying
the acceptance and isolation cuts
\begin{equation}
\begin{array}{lll}
  p_T^\ell > 10 \hbox{ GeV} \;\;,\;\; & |\eta_\ell|<2.5 \;\;,\;\; & 
  \\
  p_T^j >20  \hbox{ GeV} \;\;,\;\; & |\eta_j |<3 \;\;,\;\; &
  \\
  \Delta R_{\ell\ell}\geq 0.4 \;\;,\;\; & \Delta R_{\ell j}\geq 0.4 \;\;,\;\; & 
      \Delta R_{jj}\geq 0.4 \;\;.
\end{array}
\label{acciso4}
\end{equation}
As we can see from Table \ref{tcross4}, the SM background is dominated
by the production of $Z$ pairs, with $W^+W^- Zjj$ production
\footnote{We verified that the SM production of $W^+W^-W^+W^-jj$ has a
  negligible cross section.} a distant second.  These backgrounds can be easily removed
by requiring that
\begin{equation}
p_T^{j_{1,2}} > 50 \hbox{ GeV} \;\;\;\; \hbox{ and } \;\;\;\;
    m_{\ell^+ \ell^-} > 100 \hbox{ GeV} \; , 
\label{4lep}
\end{equation}
where $ m_{\ell^+ \ell^-}$ stands for the invariant mass of any
opposite-charge and same-flavor dileptons. As expected, the SM
background is essentially eliminated by the invariant mass
requirement. However, the signal is also considerably 
reduced. Requiring five events to establish the four-lepton 
signal, demands an integrated luminosity of 0.6 (1.7 or 4.9)
fb$^{-1}$ for a $D_4$ mass of 300 (450 or 600) GeV. Thus, although this
final state will obviously not be used for an early discovery, its
presence constitutes further evidence of the production of a
fourth-generation heavy quark.
\begin{table}[h]
\begin{center}
  \begin{tabular}{| c | c | c | } 
\toprule
   process/cuts & (\ref{acciso4}) & (\ref{acciso4}) and (\ref{4lep}) \\ 
\midrule
  signal: $m_{D_4}=300$ GeV & 59.6 & 7.87 \\
\midrule
  QCD: $m_{D_4}=300$ GeV & 58.6 & 7.52 \\
\midrule
  signal: $m_{D_4}=450$ GeV & 10.0 & 2.92 \\
\midrule
  QCD: $m_{D_4}=450$ GeV & 9.2 & 2.63 \\
\midrule
  signal: $m_{D_4}=600$ GeV & 2.4 & 1.02 \\
\midrule
  QCD: $m_{D_4}=600$ GeV & 2.1 & 0.88\\
\midrule
  $ZZjj$ & 5.0 & - \\
\midrule
  $W^+W^-Zjj$ & 0.2 & - \\
\bottomrule
  \end{tabular}
\end{center}
\caption{ Same as Table \ref{tcross3} but for the production of four charged 
  leptons.}
\label{tcross4}
\end{table}
\section{The Mechanism of Electroweak Symmetry Breaking}\label{kkgluon}

The breaking of the electroweak symmetry by the condensation of the fourth
generation relies in the strong coupling between the KK gluon and the
quarks of the fourth generation~\cite{bd}. Therefore, a conclusive test of the
scenario described in Section~\ref{effmodel} would be the detection of
the KK gluon through its  strong
coupling to the fourth-generation quarks. 
Here we briefly discuss the feasibility of such an important test.

\begin{figure}[t]
\begin{center} 
\includegraphics[width=11cm]{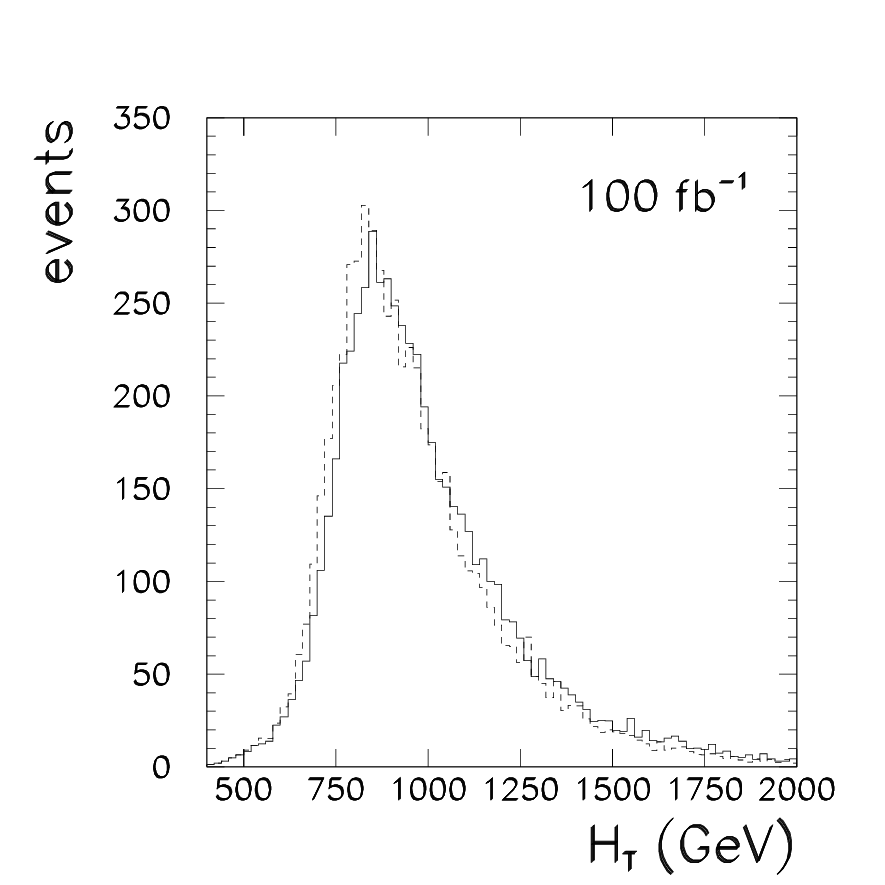}
\caption{Number of events as a function of $H_T$ for the signal with
  two same-sign leptons. We applied cuts (\ref{acciso}),
(\ref{cutjet}), and (\ref{cutlep}) and assumed an integrated
luminosity of 100 fb$^{-1}$. The dashed line corresponds to pure QCD
with $m_{D_4}=435$ GeV and the continuous one to the 5D scenario with
$m_{D_4}=450$ GeV. The total number of events is the same in both
cases.}
\label{fig4d5d}
\end{center}
\end{figure}

As mentioned in Section~\ref{effmodel}, the KK gluon is too broad for
its resonant peak to be observed. Its only effect is to give an extra 
contribution to the production of the fourth-generation quarks,
particularly $D_4$. 
However, since
the cross section is very sensitive to the $D_4$ mass, pure QCD with a
lighter $D_4$ quark can mimic the effect of the KK gluon. 
For concreteness, let us consider the signal with two
same--sign leptons and $m_{D_4}=450$ GeV; see
Section~\ref{analysis2l}. The corresponding cross section after the
cuts (\ref{acciso}), (\ref{cutjet}), and (\ref{cutlep}) is equal to
the cross section with $m_{D_4}=435$ GeV in pure QCD, {\it i.e.} a
lighter $D_4$ quark and no KK gluon. Although the distributions are
not identical, they are very similar. To distinguish them 
would require not only a very large luminosity, but also a
determination of $m_{D_4}$ with an uncertainty smaller than 15 GeV. As
an example, in Fig.~\ref{fig4d5d} we show the number of events as a
function of $H_T$ for both scenarios with an integrated luminosity of
$100fb^{-1}$. Since the shapes are indistinguishable, we conclude that
it would be very hard to detect the presence of the KK-gluon component
in this way. 
In principle, there are other tests of this scenario that could be
performed, albeit indirect. One such 
example is the detection of the KK excitations of the electroweak
gauge bosons, which would also have large
couplings to the fourth generation. We estimate the width of these
resonances to be in the range $(10-20)\%$ of their masses, making their
detection somewhat easier than in the KK gluon case.  Although the
electroweak KK modes induce only a small contribution to the
four--fermion interaction leading to the condensation, the reason
behind the large couplings is the same, i.e. the heavy fourth generation
and the massive vectors are composite states of a strongly interacting
sector with couplings $g_{SM}\lesssim g\lesssim 4\pi$. Then, the
existence of KK excitations of the electroweak gauge bosons with
large couplings to the fourth
generation would be a very important indication for the present
scenario. We leave a careful study of these signals for future work.

\section{Conclusions}\label{conclusions}
We have considered the first LHC signals of a heavy fourth-generation 
decaying preferentially to the third generation. These signals
are present in a scenario where the electroweak symmetry is broken by
the condensation of at least one of the fourth-generation
quarks~\cite{bd}. We focused on the production and decays of $D_4$,
the down quark of the fourth generation, which is assumed to be
lighter than $U_4$, the up-type fourth-generation quark. 

The existence of $D_4$  necessarily leads to
final states with  same--sign dileptons, trileptons and four charged
leptons at the LHC. These signals can be used to falsify a
model of EWSB by the condensation of the fourth generation such as the
one presented in Ref.~\cite{bd}.  We
showed that the same--sign dilepton channel, as well as the trilepton
one, should be observable at the LHC for integrated luminosities
smaller than 1~fb$^{-1}$. The non-observation of such signals
would be a 
strong constraint on this class of models, leading to many such
scenarios being ruled out. The presence of this signal, on the other
hand, would not necessarily point to the existence of a fourth
generation, which would require a lot more data than 1~fb$^{-1}$ to be
confirmed. But, to say the least, the observation of such signals 
at such low accumulated luminosities, would indicate
that the new physics is produced through 
strong interactions such as in the present scenario, but also as  in most supersymmetric
models~\cite{Baer:2008ey} among others, and not by electroweak processes.  For
instance, hundreds of fb$^{-1}$ are needed to establish the production
of heavy Majorana neutrinos~\cite{Han:2006ip} or to test the type~II
seesaw models~\cite{Perez:2008ha}. To establish the existence of 
fourth generation quarks would further require to fully reconstruct
the $D_4$ invariant mass~\cite{Holdom:2007nw}, which would take not only larger data
samples but also a good understanding of the detectors. 

Finally, to confirm the existence of a strong interaction involving
the fourth generation quarks and leading to EWSB it would be
necessary, 
in addition to the QCD-induced production of $D_4$, 
to directly observe this new strong interaction. We have shown that
even when modeled as coming from
integrating out a massive color-octet vector particle, or a KK
excitation of the gluon in AdS$_5$ models~\cite{bd}, this results in a
featureless enhancement of the $D_4$ production. In the language of
having a vector particle mediating the condensing interaction, its
width is large (of the order of its mass). 
We studied the effects of the presence of this 
KK gluon in the same--sign dilepton channel and we showed that a
very large luminosity, beyond what is foreseeable in the near future, is needed to
disentangle the QCD and KK gluon contributions. We conclude that other
strategies must be pursued in order to observe the new strong
interactions of the fourth generation. These might include the
observation of the KK excitations of the weak gauge bosons decaying
into the fourth-generation leptons, as well as the flavor-violating
effects of the KK excitations  that are only present in their
amplitude and do not suffer contamination from the SM QCD or
electroweak amplitudes. We leave their study for future
work.

\section*{Acknowledgements}

This work has been supported by Conselho Nacional de Desenvolvimento
Cient\'{\i}fico e Tecnol\'ogico (CNPq) and by Funda{c}\~ao de Amparo
\`a Pesquisa do Estado de S\~ao Paulo (FAPESP)


\end{document}